\documentclass[aps,pra,twocolumn,showpacs,amsmath,amssymb,preprintnumbers,,superscriptaddress,10pt]{revtex4-1}

\usepackage{amsmath,amssymb}
\usepackage{bm}
\usepackage[utf8]{inputenc}

\usepackage{graphicx}
\usepackage{SIunits}
\usepackage{hyphenat}
\usepackage[bookmarks=false]{hyperref}
\usepackage{color}
\usepackage[capitalise]{cleveref}

\usepackage{pifont} 
\usepackage[nomessages]{fp} 

\definecolor{pink}{RGB}{255,0,255}

\definecolor{ss_color}{rgb}{1,0,0}

\begin{document}

\title{Finite-key-size effect in a commercial plug-and-play QKD system}

\author{Poompong~Chaiwongkhot}
\email{poompong.ch@gmail.com}
\affiliation{Institute for Quantum Computing, University of Waterloo, Waterloo, ON, N2L~3G1 Canada}
\affiliation{Department of Physics and Astronomy, University of Waterloo, Waterloo, ON, N2L~3G1 Canada}

\author{Shihan~Sajeed}
\affiliation{Institute for Quantum Computing, University of Waterloo, Waterloo, ON, N2L~3G1 Canada}
\affiliation{\mbox{Department of Electrical and Computer Engineering, University of Waterloo, Waterloo, ON, N2L~3G1 Canada}}

\author{Lars Lydersen}
\affiliation{Department of Electronics Systems, Norwegian University of Science and Technology, NO-7491 Trondheim, Norway}

\author{Vadim~Makarov}
\affiliation{Department of Physics and Astronomy, University of Waterloo, Waterloo, ON, N2L~3G1 Canada}
\affiliation{Institute for Quantum Computing, University of Waterloo, Waterloo, ON, N2L~3G1 Canada}
\affiliation{\mbox{Department of Electrical and Computer Engineering, University of Waterloo, Waterloo, ON, N2L~3G1 Canada}}

\date{\today}

\begin{abstract}
A security evaluation against the finite-key-size effect was performed for a commercial plug-and-play quantum key distribution (QKD) system. We demonstrate the ability of an eavesdropper to force the system to distill key from a smaller length of sifted-key. We also derive a key-rate equation that is specific for this system. This equation provides bounds above the upper bound of secure key under finite-key-size analysis. From this equation and our experimental data, we show that the keys that have been distilled from the smaller sifted-key size fall above our bound. Thus, their security is not covered by finite-key-size analysis. Experimentally, we could consistently force the system to generate the key outside of the bound. We also test manufacturer's software update. Although all the keys after the patch fall under our bound, their security cannot be guaranteed under this analysis. Our methodology can be used for security certification and standardization of QKD systems.
\end{abstract}

\maketitle

\section{Introduction}

Quantum key distribution (QKD) systems are expected to provide unconditionally secure keys between two parties \cite{bennett1984,ekert1991,lo1999,shor2000,lutkenhaus2000,renner2005}. To fulfill that expectation, every feature, imperfection, and loophole both in theory and practice has to be taken into account. One of these features is that, with limited resources and time, a QKD system can exchange only a finite length of raw key. The knowledge of an adversary about the key is estimated by the number of errors in it~\cite{bennett1992b,brassard1994}. Since the bound on the adversary's knowledge is estimated from a finite sample, the smaller the sample is, the less accurate the estimate becomes. 
Thus, the estimated knowledge might deviate from the actual value and, if it is underestimated, the security of the secret key might be compromised. Finite-key-size analysis \cite{ben-or2005,renner2005a,renner2005b,scarani2008a,cai2009,tomamichel2012} takes these statistical deviations into account and modifies the key-rate equation accordingly.

Many of the practical QKD systems used today were developed before the finite-key-size analysis in QKD protocols became available. Although some form of finite-key-size effect has been considered in the literature since the year 2000~\cite{shor2000}, a rigorous proof was first published in 2005 and developed in the subsequent years \cite{ben-or2005,renner2005a,renner2005b,scarani2008a,cai2009,tomamichel2012}. While the finite-size analysis was not considered in the security assumptions of the early systems, the generated secret key may still be secure if the raw-key sample size is large enough to neglect the finite-size effects. However, if the sample size is smaller, the effects can no longer be neglected and an absence of the finite-key analysis may render the generated key insecure. This is the main focus of this work. We emphasize the significance of the finite-key-size effects in a practical QKD system. We also demonstrate the ability of an eavesdropper to amplify these effects by actively interfering with the transmission and forcing the system to generate secret key from a smaller sample size. In \cref{sec:experiment}, we experimentally demonstrate a simple attack that forces a commercial QKD system to use a smaller sample size. The key-rate equation for this specific system is derived in \cref{sec:derivation}. In \cref{sec:analysis}, we compare the finite-key security bounds with our experimental data. We test the system again after manufacturer's security update in \cref{sec:patch-testing}, and conclude in \cref{sec:conclusion}.

\section{Experiment}
\label{sec:experiment}

\begin{figure}
\centering
\includegraphics{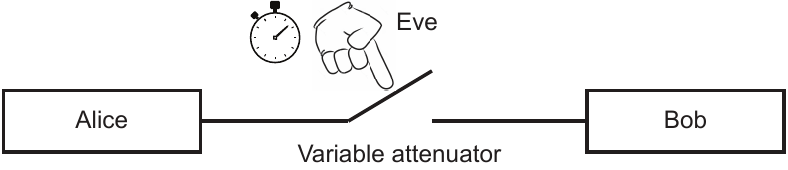}
\caption{Scheme of experiment.}
\label{fig:setup}
\end{figure}

\begin{figure*}
\includegraphics{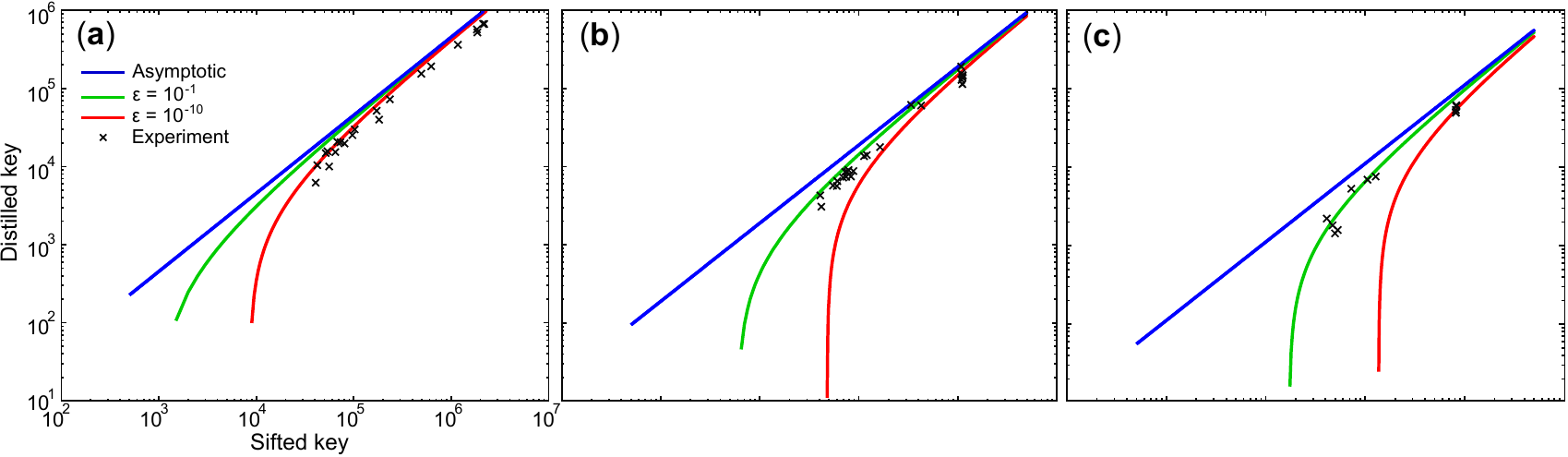}
\caption{Secret-key size versus sifted-key size. Black $\times$ are experimental results with (a) $2~\deci\bel$ line loss and $2.5\%$ error rate averaged over all sessions, (b) $3~\deci\bel$  line loss with $5.2\%$ error rate, and (c) $4~\deci\bel$ line loss with $6.2\%$ error rate. The experiment error rate in individual sessions deviated less than $\pm0.5\%$ from the average. Blue (dark grey) curve is the infinite-key bound obtained from \cref{eq:asymp}. Red (grey) curve is a finite-key-size bound with $\varepsilon = 10^{-10}$ using \cref{eq:finite}. Green (light grey) curve is the finite-key-size bound with $\varepsilon = 10^{-1}$. Secure-key bounds in each sub-figure were calculated separately according to the error rate and line loss of each experiment. Experimental points that lie above each line are considered insecure under those conditions used in each calculation. The detector's efficiency during the experiment was 0.08. The dark count rate was $2\times 10^{-5}$ counts per second for each detector.} 
\label{fig:ln-plot}
\end{figure*}

The subject of this study is the security of a plug-and-play QKD system Clavis2 produced by ID~Quantique \cite{stucki2002,idqclavis2specs}. Although updated configurations for plug-and-play systems exist \cite{zhao2010}, we have not modified the system under test and all tests were performed in the same configuration, as provided by the manufacturer. The QKD protocol under study is Bennett-Brassard 1984 (BB84) protocol without decoy states \cite{bennett1984,bennett1992b}, as implemented in Clavis2. The security of this system implemented in the manufacturer's software is based on the security analysis in Ref.~\onlinecite{niederberger2005}, which is an analysis against photon-number-splitting attack and cloning attack. The analysis in Ref.~\onlinecite{niederberger2005} neither considers the finite-key-size effects nor takes into account the lack of phase randomization in the system. It also assume that Eve cannot change detectors efficiency.

Under normal operation, the system exchanges the quantum signals until the memory buffer for the states sent by Alice is filled. This leads to the raw-key size being limited. This limit varies depending on the line loss (at higher loss fewer photons are received by Bob, and the key is smaller as our experimental data show below). Then, Alice and Bob perform post-processing: sifting, error correction, and privacy amplification \cite{bennett1992b,stucki2002,idqclavis2specs}. One of the features of Clavis2 is that the system terminates the raw-key exchange process if Bob's photon detection efficiency drops below a certain threshold, and performs a recalibration procedure for the timing alignment of detector gates \cite{idqclavis2specs,jain2011}. This timing alignment greatly affects the photon detection efficiency, is sensitive to environmental fluctuations, and needs to be restored from time to time by performing this recalibration. However the system does not discard the raw key already accumulated in the buffer (as long as it has accumulated at least $80$~kbit), and performs the post-processing from the available amount at the time of termination. Eve may take an advantage of this feature. Since the security proof of the system did not take into account the statistical deviation due to non-infinite key length, the deviation can be further amplified if the interruption for recalibration occurs early in the raw-key exchange session.

To demonstrate Eve's ability to force the system to distill from a short raw-key length, we first ran the system in a normal operation mode. The quantum channel between Alice and Bob consisted of a $2~\meter$ long optical fiber, and a variable attenuator (OZ~Optics DD-100-11-1550) was added to simulate transmission line loss of $2$, $3$ and $4~\deci\bel$ (see \cref{fig:setup}). We ran multiple sessions of key distribution. In each session, during the raw-key exchange phase, we let the system exchange quantum signals for time $\tau$, then abruptly increased the attenuation to $\approx 40~\deci\bel$. This reduced the detection rate in Bob below the threshold and forced the system to terminate the key exchange. After that, the system performed post-processing of the already exchanged raw key and reported the secret-key length for that session. At the same time, we reset the variable attenuator to the original loss value. The system then recalibrated the timing alignment, and proceeded to the next raw-key exchange session. We varied $\tau$ between $10$--$280~\second$, so that the raw-key size after termination was between the system's minimum threshold of $80$~kbit and the memory buffer limit of $1.6$--$4$~Mbit in Bob (depending on the line loss), corresponding to the leftmost and rightmost experimental points in each plot in \cref{fig:ln-plot}. We also allowed the system to complete some of the sessions naturally without Eve's intervention, which mostly resulted in the maximum key length but occasionally a shorter one. The plots show the variation of secret-key size as a function of the sifted-key size, for different transmission loss values. Note that the sifted-key size plotted is half the raw-key size. The amount of the raw key exchanged did not depend solely on $\tau$. Some sessions experienced fluctuations in transmission loss and detection rate, which caused a lower key exchange rate but not below the termination threshold. Some sessions terminated before we induced the loss, when the detection efficiency dropped below the threshold as the result of naturally occurring timing drift, without Eve's help.

In our analysis, we consider the length of secret key as a function of the sifted-key length, rather than the session time duration. For each session that produced non-zero secret key, we recorded the length of the sifted key, the number of bits disclosed in the error correction, the error rate, and the length of the secret key reported by the system. The system under test did not include finite-key-size analysis in its post-processing. Rather, the post-processing step was programmed to subtract an arbitrarily chosen amount of the key in addition to the value given by the asymptotic security analysis \cite{IDQ-priv2014}. This subtraction was done to account for any unknown effects that were not included in the system's security analysis. Prior to this study, the security of this arbitrary key subtraction has not been verified. We check this hypothesis below. Note that we consider only the case where Eve attempts to control the sifted-key-size before the post-processing. No other attack or flaw is considered in this study.

\section{Derivation of key-rate equation}
\label{sec:derivation}

For finite-key-size effects, we need to formulate the key-rate equation for this specific system. To our knowledge, there is no finite-key-size analysis that covers all assumption in this system without hardware modification \cite{zhao2007}. In this section, we use available derivation technique to find a secure key bound of the key generated by the system. We assume here that Eve does not interfere with the bright pulses sent from Bob to Alice, and assume that the phase of signal in different time slots is random. As a result, the key bound in this analysis would lie above the upper bound of secure key, which takes into account the lack of phase randomization \cite{lo2005}. Although we cannot conclude that the keys below our bound are secure, it can be used to justify that the secret keys that fall above this bound are not covered by the finite-key-size analysis. Thus, we need to assume the worst case that such keys are insecure. 

Our analysis covers the process starting with the raw-key exchange step of plug-and-play system, where Alice attenuates the laser pulses from Bob and encodes each pulse in one of the four possible phase values: $0$, $\pi/2$, $\pi$, and $3\pi/2$. Alice then sends the encoded signal back to Bob where he measures the signal in one of the two bases, and gets his raw key. They perform sifting and error correction afterwards. The system then performs privacy amplification process where the key is shortened with a universal-2 hash function to exclude Eve's information about the key. The key after this step is the secret key. Eve's information is estimated from quantum bit error rate (QBER) found during the error correction and probability of having multi-photon pulses during raw key exchange. This process allows us to use a common procedure of secret-key analysis based on Refs.~\onlinecite{ben-or2005,renner2005a,renner2005b}, which stated that, by using the universal-2 hash function as privacy amplification, a secret key $K$ of secret key probability per bit $l_K$ is $\varepsilon$-secure if the protocol is not aborted, and $l_K$ satisfies the relation
\begin{equation}
\label{eq:generic}
	\varepsilon_\text{PA} < 2^{-\frac{1}{2}(H_{min}(K|E')-l_K)}.
\end{equation}
Here $\varepsilon_\text{PA}$ is the collision probability of hash function, which is the probability of two different input strings being projected into the same string of output. $H_{min}(K|E')$ is smooth min-entropy of the system, which represents the probability of Eve guessing the key $K$ correctly using an optimal strategy, given her information about the key before privacy amplification $E'$.

The goal of this derivation is to replace the smooth min-entropy with a function of measurable parameters from the system. Since the information leakage during error correction is independent of other processes prior to that, $E'$ can be decomposed into Eve's knowledge before error correction $E$ and information leakage during error correction process $L$. By inequality of smooth entropy \cite{renner2005b}, we have  
\begin{equation}
\label{eq:KE'}
	H_{min}(K|E') \geq H_{min}(K|E'') - L -7\sqrt[]{\frac{1}{n}\log_2\frac{2}{\tilde{\varepsilon}}},
\end{equation}
where $H_{min}(K|E'')$ is the smooth min-entropy of the system before the error correction step. The last term is a statistical correction under finite-key-size regime, where $\tilde{\varepsilon}$ is the probability that Eve's information is underestimated when using smooth min-entropy \cite{scarani2008a}. The analysis in Refs.~\onlinecite{scarani2008a,tomamichel2011} gave us the bound of $H_{min}(K|E'')$ as a function of measurable parameters
\begin{equation}
\label{eq:KE}
	H_{min}(K|E'') \geq A\left(1-h\left(\frac{\tilde{E}}{A}\right)\right),
\end{equation}
where $h(x)=-x\log_2(x)-(1-x)\log_2(1-x)$ is the binary Shannon entropy, $\tilde{E}=E+\frac{1}{2}\sqrt{\{2\ln(1/\varepsilon_\text{PE})+2\ln(n+1)\}(1/n)}$ takes into account a chance that the error rate estimated from a sifted key of size $n$ in the protocol might deviate from the actual value \cite{scarani2008a}, $\varepsilon_\text{PE}$ is the probability that such deviation occurs, and $E$ is the observed error rate. The single photon detection probability $A = (p_\text{det}-p_\text{multi})/p_\text{det}$ is a correction term for weak coherent laser used to exchange the raw key in the system \cite{lutkenhaus2000}, where $p_\text{det}$ is the probability of detection and $p_\text{multi}$ is the probability of a multi-photon pulse generated by Alice \cite{idqclavis2specs}. 


Now we consider information leakage during the error correction. In theory, the minimum portion of the key with error probability $E$ that needs to be disclosed to correct all the errors is $h(E)$. Using this limit along with the finite-key-size analysis from Ref.~\onlinecite{cai2009}, we have the upper bound of information leakage during error correction  
\begin{equation}
\label{eq:EC} 
 L \leq leak_\text{EC}A + \log_2{\frac{8}{\varepsilon_\text{EC}}},
\end{equation}
where $leak_\text{EC} = f_\text{EC}h(E)$ is an estimated portion of the key disclosed during error correction. The factor $f_\text{EC} = 1.2$ is a practical efficiency of the error correction protocol \cite{brassard1994,scarani2008a}. In the system log of system under test, this value varied between $1.1$--$1.3$. The last term takes account of a failure probability $\varepsilon_\text{EC}$ that the error correction leaves non-zero number of errors \cite{cai2009}. This can occur, for example, owing to a non-zero probability of at least one parity check block containing an even number of error bits in every iteration of CASCADE error-correction code and the following parity check rounds in Clavis2 \cite{idqclavis2specs}.

Since the experimental results are the secret key size as a function of the sifted key length $n$, we need a secure key bound $l=nl_K$. Substituting \cref{eq:KE',eq:KE,eq:EC} into \cref{eq:generic}, taking the logarithm, then multiplying by $n$ on both sides, we obtain 
\begin{equation} \label{eq:finite}
\begin{aligned}
l \quad \leq & \quad n A\left(1-h\left(\frac{\tilde{E}}{A}\right)\right)-n \ leak_\text{EC}\\
&-7n\,\sqrt[]{\frac{1}{n}\log_2\frac{2}{\tilde{\varepsilon}}}-2\log_2\frac{1}{\varepsilon_\text{PA}}-\log_2\frac{2}{\varepsilon_\text{EC}},
\end{aligned}
\end{equation}  
with security parameter $\varepsilon = \varepsilon_\text{PE} + \tilde{\varepsilon} + \varepsilon_\text{PA} + \varepsilon_\text{EC}$ \cite{renner2005b,scarani2008a,cai2009,tomamichel2012}. Since secret-key-rate analyses under collective and coherent attack on non-decoy state BB84 are equivalent \cite{renner2005,kraus2005}, the present analysis also covers coherent attack, which is the most general form of attacks on QKD system.

The asymptotic key-rate equation for this specific system can be derived in the same way, but without considering statistical deviation due to finite-size effects. The asymptotic key-rate is
\begin{equation} \label{eq:asymp}
l_\infty \quad \leq \quad n A\left(1-h\left(\frac{{E}}{A}\right)\right)-n\ leak_\text{EC}.
\end{equation}
 
\section{Security verification}
\label{sec:analysis}
To verify the security of the secret key, we compare the experimental result with the bound of the secret key under the asymptotic assumption and finite-key-size analysis. For the asymptotic case, we use \cref{eq:asymp} as the secure key bound. For finite-key-size effect, we use a numerical optimization to find a combination of security parameters ($\varepsilon_\text{PE}, \tilde{\varepsilon},\varepsilon_\text{PA}$, and $\varepsilon_\text{EC}$) that maximizes the key length in \cref{eq:finite}. The observed error rate $E$ is an average of the error rates reported by the system after each key distillation at a given transmission loss. The term $A$ is calculated assuming the Poisson distribution with a mean photon number per pulse $\mu = 0.2$ sent by Alice. The value of $\mu$ varied between $0.2$--$0.4$ in the experiment, however the lowest value gives the highest bound for the secret key rate. We thus obtain  bounds of secure key length, plotted in \cref{fig:ln-plot}. Above each curve lies the zone where the security of the key is not covered by finite-key-size analysis.

The experimental secret key sizes, denoted by black $\times$, always satisfy the security criteria for the asymptotic assumption. When the size of the input sifted key is large, the key-rate bounds with and without finite-size assumption lie very close to each other (see~\cref{fig:ln-plot}). This might put the experimentally distilled key size below the finite-key-size bound, i.e., on the safe side. However, when the sifted key size is reduced, the key-rate bounds with and without finite-key assumption diverge significantly. Higher loss results in higher divergence. A fraction of the experimental results falls outside the secure zone for the finite-key-size analysis with values of $\varepsilon$ up to $10^{-1}$. The latter value means there is a $10\%$ chance that the information of the key generated under this condition might be leaked to Eve. In practice, the security parameter $\varepsilon$ can be picked to be of the same order as the probability of major natural disasters such as a serious earthquake, nearby volcanic eruption or nuclear power plant meltdown \cite{renner-priv2014}. If such disaster happened, it is most likely that the security of the key would not matter anymore. For example, the probability of a nuclear power plant meltdown is $10^{-4}$ per year, according to the Nuclear Regulatory Commission \cite{nrc2004}. If our QKD machine generates two keys every minute or approximately $10^6$ keys a year, one might pick $\varepsilon=10^{-10}$ so that the probability that at least one key leaks to Eve is of the same order as such disasters \cite{renner-priv2014}. However, our experiment shows that Eve can consistently induce a much higher risk probability of key leakage. She can do this by applying our channel interruption technique for BB84 protocol at channel loss values $> 2~\deci\bel$ (or line distances longer than about $12~\kilo\meter$, given typical fiber loss value of $0.17~\deci\bel\per\kilo\meter$).

\section{Testing manufacturer's patch}
\label{sec:patch-testing}

In the middle of our study in 2014, ID~Quantique released a software update for Clavis2. After the update, the system accumulates the raw key over multiple key exchange sessions, and performs post-processing only when the sifted-key size reaches a threshold of about $2$~Mbit.

\begin{figure}
\centering
\includegraphics{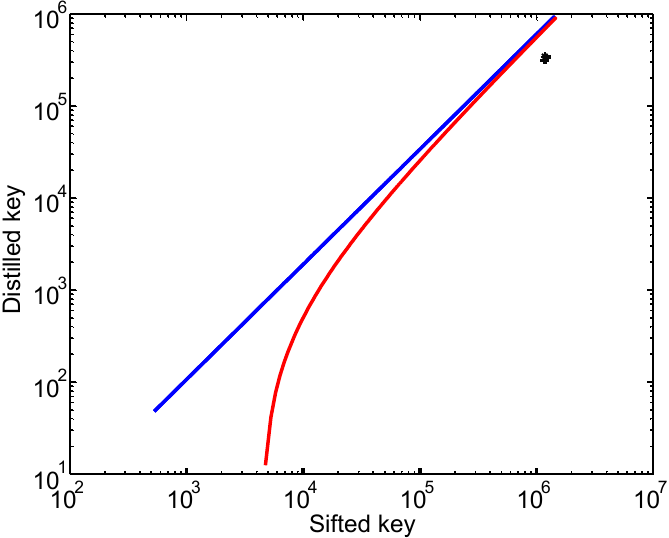}
\caption{Experimental result with new software. The line loss was $3~\deci\bel$ and error rate was $1\%$. Blue (dark grey) curve is the asymptotic key bound given by \cref{eq:asymp}. Red (grey) curve is the finite-key-size bound for $\varepsilon = 10^{-10}$ given by \cref{eq:finite}. A group of $\times$ presents results of 8 key distillations.}
\label{fig:ln-plot-patch}
\end{figure}

We have repeated our experiment and recalculated our plot using the new parameters acquired from the updated system. The result shows that the secret key is within the secure bound of $\varepsilon = 10^{-10}$ (see \cref{fig:ln-plot-patch}). Regardless of our channel interruptions, we observed that the system has retained the raw key exchanged before termination of each raw-key exchange session, and accumulated it until the size reached about $2$~Mbit before proceeding to the distillation. This behavior is clearly visible in the system log and confirmed by the manufacturer \cite{IDQ-priv2014}.

\section{Conclusion}
\label{sec:conclusion}

In this work, we have done a security evaluation of the finite-key-size effect for Clavis2 system that included derivation of the specific key-rate equation, developing a testing methodology, using it to test the system's security against finite-key-size effects, and testing the manufacturer's patch. Although rigorous security proofs with finite-key-size assumptions were abundant in the literature during the start of this work, they were not assembled together into a key-rate equation suitable for the system under test. Our work has assembled the components of the key-rate equation, verified the assumptions, and put them together into the form of \cref{eq:finite}. However, under our assumptions, the equation does not give the upper bound to evaluate the security of the secret key. Using our result, we can only verify that the keys that fall above the bound are not secure under finite-key-size analysis. 

We have shown that by dynamically controlling the channel loss, Eve can force the system to distill key from a shorter sifted-key length to bring the finite-key-effects into play. Using our derived key-rate equation, \cref{eq:finite}, we have shown that key distilled from a sufficiently small length of sifted-key is not guaranteed to be secure, even with the manufacturer's added post-processing step of secret-key subtraction. We have also investigated the security update from ID~Quantique, and found that all experimental results fall under the bound in this study. Unfortunately the security of the key against this attack cannot be concluded from this result. Our study only covers statistical evidence from the system against a theoretical bound. An explicit attack that exploits this effect is still open for future study. 

Our investigation highlights the significance and importance of finite-key-size analysis in the implementations of QKD, especially in commercial systems. Our method of attack can be used as basis of a testing methodology for security certification. It should be incorporated in the standardization of QKD, which is the next step this technology field faces \cite{langer2009}.

We responsibly disclosed to ID~Quantique partial results of this investigation before the 2014 patch. Publication has been delayed in order to give the company enough time for patch deployment.

\bigskip

\noindent {\bf Acknowledgments.} We thank the anonymous referee for his/her insightful and thorough criticism that led to improvement of the theory in this study. We thank M.~Curty, N.~L{\"u}t\-ken\-haus, R.~Renner, and J.~Skaar for discussions. We thank ID~Quantique for cooperation, technical assistance, and providing us the QKD hardware. This work was supported by Industry Canada, NSERC (programs Discovery and CryptoWorks21), CFI and Ontario MRI. P.C.\ acknowledges support by Thai DPST scholarship.

\medskip

\noindent {\bf Competing interests.} The authors have a research collaboration on security analysis of QKD products with ID~Quantique. No other competing interests are declared.

\bibliography{bibtex_library}

\begin{thebibliography}{27}%
\makeatletter
\providecommand \@ifxundefined [1]{%
 \@ifx{#1\undefined}
}%
\providecommand \@ifnum [1]{%
 \ifnum #1\expandafter \@firstoftwo
 \else \expandafter \@secondoftwo
 \fi
}%
\providecommand \@ifx [1]{%
 \ifx #1\expandafter \@firstoftwo
 \else \expandafter \@secondoftwo
 \fi
}%
\providecommand \natexlab [1]{#1}%
\providecommand \enquote  [1]{``#1''}%
\providecommand \bibnamefont  [1]{#1}%
\providecommand \bibfnamefont [1]{#1}%
\providecommand \citenamefont [1]{#1}%
\providecommand \href@noop [0]{\@secondoftwo}%
\providecommand \href [0]{\begingroup \@sanitize@url \@href}%
\providecommand \@href[1]{\@@startlink{#1}\@@href}%
\providecommand \@@href[1]{\endgroup#1\@@endlink}%
\providecommand \@sanitize@url [0]{\catcode `\\12\catcode `\$12\catcode
  `\&12\catcode `\#12\catcode `\^12\catcode `\_12\catcode `\%12\relax}%
\providecommand \@@startlink[1]{}%
\providecommand \@@endlink[0]{}%
\providecommand \url  [0]{\begingroup\@sanitize@url \@url }%
\providecommand \@url [1]{\endgroup\@href {#1}{\urlprefix }}%
\providecommand \urlprefix  [0]{URL }%
\providecommand \Eprint [0]{\href }%
\providecommand \doibase [0]{http://dx.doi.org/}%
\providecommand \selectlanguage [0]{\@gobble}%
\providecommand \bibinfo  [0]{\@secondoftwo}%
\providecommand \bibfield  [0]{\@secondoftwo}%
\providecommand \translation [1]{[#1]}%
\providecommand \BibitemOpen [0]{}%
\providecommand \bibitemStop [0]{}%
\providecommand \bibitemNoStop [0]{.\EOS\space}%
\providecommand \EOS [0]{\spacefactor3000\relax}%
\providecommand \BibitemShut  [1]{\csname bibitem#1\endcsname}%
\let\auto@bib@innerbib\@empty
\bibitem [{\citenamefont {Bennett}\ and\ \citenamefont
  {Brassard}(1984)}]{bennett1984}%
  \BibitemOpen
  \bibfield  {author} {\bibinfo {author} {\bibfnamefont {C.~H.}\ \bibnamefont
  {Bennett}}\ and\ \bibinfo {author} {\bibfnamefont {G.}~\bibnamefont
  {Brassard}},\ }in\ \href@noop {} {\emph {\bibinfo {booktitle} {Proc. IEEE
  International Conference on Computers, Systems, and Signal Processing
  (Bangalore, India)}}}\ (\bibinfo  {publisher} {IEEE Press},\ \bibinfo
  {address} {New York},\ \bibinfo {year} {1984})\ pp.\ \bibinfo {pages}
  {175--179}\BibitemShut {NoStop}%
\bibitem [{\citenamefont {Ekert}(1991)}]{ekert1991}%
  \BibitemOpen
  \bibfield  {author} {\bibinfo {author} {\bibfnamefont {A.~K.}\ \bibnamefont
  {Ekert}},\ }\href {\doibase 10.1103/PhysRevLett.67.661} {\bibfield  {journal}
  {\bibinfo  {journal} {Phys. Rev. Lett.}\ }\textbf {\bibinfo {volume} {67}},\
  \bibinfo {pages} {661} (\bibinfo {year} {1991})}\BibitemShut {NoStop}%
\bibitem [{\citenamefont {Lo}\ and\ \citenamefont {Chau}(1999)}]{lo1999}%
  \BibitemOpen
  \bibfield  {author} {\bibinfo {author} {\bibfnamefont {H.-K.}\ \bibnamefont
  {Lo}}\ and\ \bibinfo {author} {\bibfnamefont {H.~F.}\ \bibnamefont {Chau}},\
  }\href {\doibase 10.1126/science.283.5410.2050} {\bibfield  {journal}
  {\bibinfo  {journal} {Science}\ }\textbf {\bibinfo {volume} {283}},\ \bibinfo
  {pages} {2050} (\bibinfo {year} {1999})}\BibitemShut {NoStop}%
\bibitem [{\citenamefont {Shor}\ and\ \citenamefont
  {Preskill}(2000)}]{shor2000}%
  \BibitemOpen
  \bibfield  {author} {\bibinfo {author} {\bibfnamefont {P.~W.}\ \bibnamefont
  {Shor}}\ and\ \bibinfo {author} {\bibfnamefont {J.}~\bibnamefont
  {Preskill}},\ }\href {\doibase 10.1103/PhysRevLett.85.441} {\bibfield
  {journal} {\bibinfo  {journal} {Phys. Rev. Lett.}\ }\textbf {\bibinfo
  {volume} {85}},\ \bibinfo {pages} {441} (\bibinfo {year} {2000})}\BibitemShut
  {NoStop}%
\bibitem [{\citenamefont {L\"utkenhaus}(2000)}]{lutkenhaus2000}%
  \BibitemOpen
  \bibfield  {author} {\bibinfo {author} {\bibfnamefont {N.}~\bibnamefont
  {L\"utkenhaus}},\ }\href {\doibase 10.1103/PhysRevA.61.052304} {\bibfield
  {journal} {\bibinfo  {journal} {Phys. Rev. A}\ }\textbf {\bibinfo {volume}
  {61}},\ \bibinfo {pages} {052304} (\bibinfo {year} {2000})}\BibitemShut
  {NoStop}%
\bibitem [{\citenamefont {Renner}\ \emph {et~al.}(2005)\citenamefont {Renner},
  \citenamefont {Gisin},\ and\ \citenamefont {Kraus}}]{renner2005}%
  \BibitemOpen
  \bibfield  {author} {\bibinfo {author} {\bibfnamefont {R.}~\bibnamefont
  {Renner}}, \bibinfo {author} {\bibfnamefont {N.}~\bibnamefont {Gisin}}, \
  and\ \bibinfo {author} {\bibfnamefont {B.}~\bibnamefont {Kraus}},\ }\href
  {\doibase 10.1103/PhysRevA.72.012332} {\bibfield  {journal} {\bibinfo
  {journal} {Phys. Rev. A}\ }\textbf {\bibinfo {volume} {72}},\ \bibinfo
  {pages} {012332} (\bibinfo {year} {2005})}\BibitemShut {NoStop}%
\bibitem [{\citenamefont {Bennett}\ \emph {et~al.}(1992)\citenamefont
  {Bennett}, \citenamefont {Bessette}, \citenamefont {Salvail}, \citenamefont
  {Brassard},\ and\ \citenamefont {Smolin}}]{bennett1992b}%
  \BibitemOpen
  \bibfield  {author} {\bibinfo {author} {\bibfnamefont {C.~H.}\ \bibnamefont
  {Bennett}}, \bibinfo {author} {\bibfnamefont {F.}~\bibnamefont {Bessette}},
  \bibinfo {author} {\bibfnamefont {L.}~\bibnamefont {Salvail}}, \bibinfo
  {author} {\bibfnamefont {G.}~\bibnamefont {Brassard}}, \ and\ \bibinfo
  {author} {\bibfnamefont {J.}~\bibnamefont {Smolin}},\ }\href {\doibase
  10.1007/bf00191318} {\bibfield  {journal} {\bibinfo  {journal} {J.
  Cryptology}\ }\textbf {\bibinfo {volume} {5}},\ \bibinfo {pages} {3}
  (\bibinfo {year} {1992})}\BibitemShut {NoStop}%
\bibitem [{\citenamefont {Brassard}\ and\ \citenamefont
  {Salvail}(1994)}]{brassard1994}%
  \BibitemOpen
  \bibfield  {author} {\bibinfo {author} {\bibfnamefont {G.}~\bibnamefont
  {Brassard}}\ and\ \bibinfo {author} {\bibfnamefont {L.}~\bibnamefont
  {Salvail}},\ }\href {\doibase 10.1007/3-540-48285-7_35} {\bibfield  {journal}
  {\bibinfo  {journal} {Lect. Notes Comp. Sci.}\ }\textbf {\bibinfo {volume}
  {765}},\ \bibinfo {pages} {410} (\bibinfo {year} {1994})}\BibitemShut
  {NoStop}%
\bibitem [{\citenamefont {Ben-Or}\ \emph {et~al.}(2005)\citenamefont {Ben-Or},
  \citenamefont {Horodecki}, \citenamefont {Leung}, \citenamefont {Mayers},\
  and\ \citenamefont {Oppenheim}}]{ben-or2005}%
  \BibitemOpen
  \bibfield  {author} {\bibinfo {author} {\bibfnamefont {M.}~\bibnamefont
  {Ben-Or}}, \bibinfo {author} {\bibfnamefont {M.}~\bibnamefont {Horodecki}},
  \bibinfo {author} {\bibfnamefont {D.}~\bibnamefont {Leung}}, \bibinfo
  {author} {\bibfnamefont {D.}~\bibnamefont {Mayers}}, \ and\ \bibinfo {author}
  {\bibfnamefont {J.}~\bibnamefont {Oppenheim}},\ }\href {\doibase
  10.1007/978-3-540-30576-7_21} {\bibfield  {journal} {\bibinfo  {journal}
  {Lect. Notes Comp. Sci.}\ ,\ \bibinfo {pages} {386}} (\bibinfo {year}
  {2005})}\BibitemShut {NoStop}%
\bibitem [{\citenamefont {Renner}\ and\ \citenamefont
  {Koenig}(2005)}]{renner2005a}%
  \BibitemOpen
  \bibfield  {author} {\bibinfo {author} {\bibfnamefont {R.}~\bibnamefont
  {Renner}}\ and\ \bibinfo {author} {\bibfnamefont {R.}~\bibnamefont
  {Koenig}},\ }\href {\doibase 10.1007/978-3-540-30576-7_22} {\ \bibinfo
  {series} {Lect. Notes Comp. Sci.},\ \textbf {\bibinfo {volume} {3378}},\
  \bibinfo {pages} {407} (\bibinfo {year} {2005})}\BibitemShut {NoStop}%
\bibitem [{\citenamefont {Renner}(2005)}]{renner2005b}%
  \BibitemOpen
  \bibfield  {author} {\bibinfo {author} {\bibfnamefont {R.}~\bibnamefont
  {Renner}},\ }\emph {\bibinfo {title} {Security of quantum key
  distribution}},\ \href {\doibase 10.3929/ethz-a-005115027} {Ph.D. thesis}
  (\bibinfo {year} {2005})\BibitemShut {NoStop}%
\bibitem [{\citenamefont {Scarani}\ and\ \citenamefont
  {Renner}(2008)}]{scarani2008a}%
  \BibitemOpen
  \bibfield  {author} {\bibinfo {author} {\bibfnamefont {V.}~\bibnamefont
  {Scarani}}\ and\ \bibinfo {author} {\bibfnamefont {R.}~\bibnamefont
  {Renner}},\ }\href {\doibase 10.1103/PhysRevLett.100.200501} {\bibfield
  {journal} {\bibinfo  {journal} {Phys. Rev. Lett.}\ }\textbf {\bibinfo
  {volume} {100}},\ \bibinfo {pages} {200501} (\bibinfo {year}
  {2008})}\BibitemShut {NoStop}%
\bibitem [{\citenamefont {Cai}\ and\ \citenamefont {Scarani}(2009)}]{cai2009}%
  \BibitemOpen
  \bibfield  {author} {\bibinfo {author} {\bibfnamefont {R.~Y.~Q.}\
  \bibnamefont {Cai}}\ and\ \bibinfo {author} {\bibfnamefont {V.}~\bibnamefont
  {Scarani}},\ }\href {\doibase 10.1088/1367-2630/11/4/045024} {\bibfield
  {journal} {\bibinfo  {journal} {New J. Phys.}\ }\textbf {\bibinfo {volume}
  {11}},\ \bibinfo {pages} {045024} (\bibinfo {year} {2009})}\BibitemShut
  {NoStop}%
\bibitem [{\citenamefont {Tomamichel}\ \emph {et~al.}(2012)\citenamefont
  {Tomamichel}, \citenamefont {Lim}, \citenamefont {Gisin},\ and\ \citenamefont
  {Renner}}]{tomamichel2012}%
  \BibitemOpen
  \bibfield  {author} {\bibinfo {author} {\bibfnamefont {M.}~\bibnamefont
  {Tomamichel}}, \bibinfo {author} {\bibfnamefont {C.~C.~W.}\ \bibnamefont
  {Lim}}, \bibinfo {author} {\bibfnamefont {N.}~\bibnamefont {Gisin}}, \ and\
  \bibinfo {author} {\bibfnamefont {R.}~\bibnamefont {Renner}},\ }\href
  {\doibase 10.1038/ncomms1631} {\bibfield  {journal} {\bibinfo  {journal}
  {Nat. Commun.}\ }\textbf {\bibinfo {volume} {3}},\ \bibinfo {pages} {634}
  (\bibinfo {year} {2012})}\BibitemShut {NoStop}%
\bibitem [{\citenamefont {Stucki}\ \emph {et~al.}(2002)\citenamefont {Stucki},
  \citenamefont {Gisin}, \citenamefont {Guinnard}, \citenamefont {Ribordy},\
  and\ \citenamefont {Zbinden}}]{stucki2002}%
  \BibitemOpen
  \bibfield  {author} {\bibinfo {author} {\bibfnamefont {D.}~\bibnamefont
  {Stucki}}, \bibinfo {author} {\bibfnamefont {N.}~\bibnamefont {Gisin}},
  \bibinfo {author} {\bibfnamefont {O.}~\bibnamefont {Guinnard}}, \bibinfo
  {author} {\bibfnamefont {G.}~\bibnamefont {Ribordy}}, \ and\ \bibinfo
  {author} {\bibfnamefont {H.}~\bibnamefont {Zbinden}},\ }\href {\doibase
  10.1088/1367-2630/4/1/341} {\bibfield  {journal} {\bibinfo  {journal} {New J.
  Phys.}\ }\textbf {\bibinfo {volume} {4}},\ \bibinfo {pages} {41} (\bibinfo
  {year} {2002})}\BibitemShut {NoStop}%
\bibitem [{idq()}]{idqclavis2specs}%
  \BibitemOpen
  \href@noop {} {}\bibinfo {note} {{C}lavis2 specification sheet,
  \url{http://www.idquantique.com/images/stories/PDF/clavis2-quantum-key-distribution/clavis2-specs.pdf},
  visited 4 July 2014}\BibitemShut {NoStop}%
\bibitem [{\citenamefont {Zhao}\ \emph {et~al.}(2010)\citenamefont {Zhao},
  \citenamefont {Qi}, \citenamefont {Lo},\ and\ \citenamefont
  {Qian}}]{zhao2010}%
  \BibitemOpen
  \bibfield  {author} {\bibinfo {author} {\bibfnamefont {Y.}~\bibnamefont
  {Zhao}}, \bibinfo {author} {\bibfnamefont {B.}~\bibnamefont {Qi}}, \bibinfo
  {author} {\bibfnamefont {H.-K.}\ \bibnamefont {Lo}}, \ and\ \bibinfo {author}
  {\bibfnamefont {L.}~\bibnamefont {Qian}},\ }\href {\doibase
  10.1088/1367-2630/12/2/023024} {\bibfield  {journal} {\bibinfo  {journal}
  {New J. Phys.}\ }\textbf {\bibinfo {volume} {12}},\ \bibinfo {pages} {023024}
  (\bibinfo {year} {2010})}\BibitemShut {NoStop}%
\bibitem [{\citenamefont {Niederberger}\ \emph {et~al.}(2005)\citenamefont
  {Niederberger}, \citenamefont {Scarani},\ and\ \citenamefont
  {Gisin}}]{niederberger2005}%
  \BibitemOpen
  \bibfield  {author} {\bibinfo {author} {\bibfnamefont {A.}~\bibnamefont
  {Niederberger}}, \bibinfo {author} {\bibfnamefont {V.}~\bibnamefont
  {Scarani}}, \ and\ \bibinfo {author} {\bibfnamefont {N.}~\bibnamefont
  {Gisin}},\ }\href {\doibase 10.1103/PhysRevA.71.042316} {\bibfield  {journal}
  {\bibinfo  {journal} {Phys. Rev. A}\ }\textbf {\bibinfo {volume} {71}},\
  \bibinfo {pages} {042316} (\bibinfo {year} {2005})}\BibitemShut {NoStop}%
\bibitem [{\citenamefont {Jain}\ \emph {et~al.}(2011)\citenamefont {Jain},
  \citenamefont {Wittmann}, \citenamefont {Lydersen}, \citenamefont {Wiechers},
  \citenamefont {Elser}, \citenamefont {Marquardt}, \citenamefont {Makarov},\
  and\ \citenamefont {Leuchs}}]{jain2011}%
  \BibitemOpen
  \bibfield  {author} {\bibinfo {author} {\bibfnamefont {N.}~\bibnamefont
  {Jain}}, \bibinfo {author} {\bibfnamefont {C.}~\bibnamefont {Wittmann}},
  \bibinfo {author} {\bibfnamefont {L.}~\bibnamefont {Lydersen}}, \bibinfo
  {author} {\bibfnamefont {C.}~\bibnamefont {Wiechers}}, \bibinfo {author}
  {\bibfnamefont {D.}~\bibnamefont {Elser}}, \bibinfo {author} {\bibfnamefont
  {C.}~\bibnamefont {Marquardt}}, \bibinfo {author} {\bibfnamefont
  {V.}~\bibnamefont {Makarov}}, \ and\ \bibinfo {author} {\bibfnamefont
  {G.}~\bibnamefont {Leuchs}},\ }\href {\doibase
  10.1103/PhysRevLett.107.110501} {\bibfield  {journal} {\bibinfo  {journal}
  {Phys. Rev. Lett.}\ }\textbf {\bibinfo {volume} {107}},\ \bibinfo {pages}
  {110501} (\bibinfo {year} {2011})}\BibitemShut {NoStop}%
\bibitem [{IDQ()}]{IDQ-priv2014}%
  \BibitemOpen
  \href@noop {} {}\bibinfo {note} {ID Quantique, private communication
  (2014).}\BibitemShut {Stop}%
\bibitem [{\citenamefont {Zhao}\ \emph {et~al.}(2007)\citenamefont {Zhao},
  \citenamefont {Qi},\ and\ \citenamefont {Lo}}]{zhao2007}%
  \BibitemOpen
  \bibfield  {author} {\bibinfo {author} {\bibfnamefont {Y.}~\bibnamefont
  {Zhao}}, \bibinfo {author} {\bibfnamefont {B.}~\bibnamefont {Qi}}, \ and\
  \bibinfo {author} {\bibfnamefont {H.-K.}\ \bibnamefont {Lo}},\ }\href
  {\doibase 10.1063/1.2432296} {\bibfield  {journal} {\bibinfo  {journal}
  {Appl. Phys. Lett.}\ }\textbf {\bibinfo {volume} {90}},\ \bibinfo {eid}
  {044106} (\bibinfo {year} {2007})}\BibitemShut {NoStop}%
\bibitem [{\citenamefont {Lo}\ \emph {et~al.}(2005)\citenamefont {Lo},
  \citenamefont {Ma},\ and\ \citenamefont {Chen}}]{lo2005}%
  \BibitemOpen
  \bibfield  {author} {\bibinfo {author} {\bibfnamefont {H.-K.}\ \bibnamefont
  {Lo}}, \bibinfo {author} {\bibfnamefont {X.}~\bibnamefont {Ma}}, \ and\
  \bibinfo {author} {\bibfnamefont {K.}~\bibnamefont {Chen}},\ }\href {\doibase
  10.1103/PhysRevLett.94.230504} {\bibfield  {journal} {\bibinfo  {journal}
  {Phys. Rev. Lett.}\ }\textbf {\bibinfo {volume} {94}},\ \bibinfo {pages}
  {230504} (\bibinfo {year} {2005})}\BibitemShut {NoStop}%
\bibitem [{\citenamefont {Tomamichel}\ and\ \citenamefont
  {Renner}(2011)}]{tomamichel2011}%
  \BibitemOpen
  \bibfield  {author} {\bibinfo {author} {\bibfnamefont {M.}~\bibnamefont
  {Tomamichel}}\ and\ \bibinfo {author} {\bibfnamefont {R.}~\bibnamefont
  {Renner}},\ }\href {\doibase 10.1103/PhysRevLett.106.110506} {\bibfield
  {journal} {\bibinfo  {journal} {Phys. Rev. Lett.}\ }\textbf {\bibinfo
  {volume} {106}},\ \bibinfo {pages} {110506} (\bibinfo {year}
  {2011})}\BibitemShut {NoStop}%
\bibitem [{\citenamefont {Kraus}\ \emph {et~al.}(2005)\citenamefont {Kraus},
  \citenamefont {Gisin},\ and\ \citenamefont {Renner}}]{kraus2005}%
  \BibitemOpen
  \bibfield  {author} {\bibinfo {author} {\bibfnamefont {B.}~\bibnamefont
  {Kraus}}, \bibinfo {author} {\bibfnamefont {N.}~\bibnamefont {Gisin}}, \ and\
  \bibinfo {author} {\bibfnamefont {R.}~\bibnamefont {Renner}},\ }\href
  {\doibase 10.1103/PhysRevLett.95.080501} {\bibfield  {journal} {\bibinfo
  {journal} {Phys. Rev. Lett.}\ }\textbf {\bibinfo {volume} {95}},\ \bibinfo
  {pages} {080501} (\bibinfo {year} {2005})}\BibitemShut {NoStop}%
\bibitem [{ren()}]{renner-priv2014}%
  \BibitemOpen
  \href@noop {} {}\bibinfo {note} {R.~Renner, private communication and
  lectures (2014).}\BibitemShut {Stop}%
\bibitem [{\citenamefont {{Office of Nuclear Regulatory
  Research}}()}]{nrc2004}%
  \BibitemOpen
  \bibfield  {author} {\bibinfo {author} {\bibnamefont {{Office of Nuclear
  Regulatory Research}}},\ }\href@noop {} {\enquote {\bibinfo {title}
  {Regulatory analysis guidelines of the {U.S.}\ {N}uclear {R}egulatory
  {C}ommission},}\ }\bibinfo {note} {{NUREG/BR-0058,} {R}ev.~4 (2004),
  \url{http://www.nrc.gov/reading-rm/doc-collections/nuregs/brochures/br0058/br0058r4.pdf}.}\BibitemShut
  {Stop}%
\bibitem [{\citenamefont {Länger}\ and\ \citenamefont
  {Lenhart}(2009)}]{langer2009}%
  \BibitemOpen
  \bibfield  {author} {\bibinfo {author} {\bibfnamefont {T.}~\bibnamefont
  {Länger}}\ and\ \bibinfo {author} {\bibfnamefont {G.}~\bibnamefont
  {Lenhart}},\ }\href {\doibase 10.1088/1367-2630/11/5/055051} {\bibfield
  {journal} {\bibinfo  {journal} {New J. Phys.}\ }\textbf {\bibinfo {volume}
  {11}},\ \bibinfo {pages} {055051} (\bibinfo {year} {2009})}\BibitemShut
  {NoStop}%
\end{thebibliography}%

\end{document}